\documentclass[prb,preprint]{revtex4}

\usepackage{graphicx}
\usepackage{amsmath}

\usepackage{epstopdf}

\def\beq{\begin{equation}}
\def\eeq{\end{equation}}

\begin{document}
\title{The two-body problem of ultra-cold atoms in a harmonic trap}
\author{Patrick Shea}
\affiliation{Department of Physics and Atmospheric Science, Dalhousie 
University, Halifax, Nova Scotia, Canada B3H 3J5}

\author{Brandon P. van Zyl}
\affiliation{Department of Physics, St.\ Francis Xavier University, 
Antigonish, Nova Scotia, Canada B2G 2W5}

\author{Rajat K. Bhaduri}
\affiliation{Department of Physics and Astronomy,
McMaster University, Hamilton, Ontario, Canada L8S 4M1}


\begin{abstract}
We consider two bosonic atoms interacting 
with a short-range
potential and trapped in a spherically symmetric harmonic
oscillator. The problem is exactly solvable and is relevant for the
study of ultra-cold atoms. We show that the energy spectrum is 
universal, irrespective of the shape of the interaction potential,
provided its range is much smaller than the oscillator length. 
\end{abstract}

\maketitle

\section{Introduction}
From the earliest studies of the dynamics of our solar system 
to the first attempts to apply quantum mechanics to the simplest atoms
(hydrogen and helium), few-body systems have played a pivotal role. 
In particular, the problem of two particles interacting via a 
central force law is familiar to 
students of physics. After elimination of the center-of-mass
motion the two-body
problem is reduced to a one-body problem in the relative co-ordinate. The well-known two-body
gravitational and the Coulomb problems admit exact analytical
solutions, which can be tested by experiments.

The few-body problem is still of great interest,
especially in view of recent experimental advancements in the
area of ultra-cold atoms.\cite{greiner,stof} 
In these experiments neutral atoms are first
cooled to nano-Kelvin temperatures and then confined
in an optical lattice formed by standing wave laser beams, where there are
as few as two-and three atoms per lattice site. 
Typically, the lattice sites confining the atoms may be regarded as 
harmonic oscillator potentials with a trapping frequency
$\nu$ of a few tens of cycles per second. The coupling between the
harmonic oscillator wells (that is, between the lattice sites) may be controlled by the
intensity of the laser beams. In the limit of low tunneling, the
individual wells may be regarded as independent three-dimensional (3D)
harmonic oscillators. 
The strength of the pair-wise interaction between the atoms on each site 
may be fine-tuned by sweeping through a static magnetic field using a Feshbach 
resonance.\cite{feshbach,ino,tannoudji} At nano-Kelvin temperatures
quantum effects dominate, and we
thus have what amounts to a non-relativistic quantum few-body system
whose inter-particle interactions can be controlled at will.
This achievement is truly remarkable, and such ultra-cold matter
experiments can serve as ideal laboratories to investigate
the quantum few-body problem. 

The experimental group of St\"oferle {\it et al.}\cite{stof} 
has trapped two $^{40}$K atoms 
in a 3D harmonic oscillator well, and by tuning near 
the Feshbach 
resonance, can make the inter-particle interaction
strong enough to form diatomic molecules. Because the two atoms
(each of which behaves as a fermion\cite{note1}) 
occupy distinct hyperfine states,
they may still interact in the relative $s$-state. Subsequently, the energy of the state is deduced by dissociating the molecule using an rf-pulse 
and measuring the binding energy of the molecule. The binding energy
as a function of the strength of the interaction (which is related to the
$s$-wave scattering length) is mapped out by
by sweeping a magnetic field using a Feshbach resonance. 
In these experiments the atoms form only loosely bound pairs so that the interaction range
may be taken to be very small compared to the size
of the two-body system. This system is then an experimental realization of the exactly solvable problem of 
two harmonically trapped particles interacting via a zero-range
pair-wise potential at zero temperature.

In this article we present the exact solution to this quantum
two-body problem, 
assuming
only a knowledge of quantum mechanics at the
senior undergraduate or beginning graduate student level. 
Remarkably, this ostensibly simple problem has only been investigated
within the last 10 years,\cite{busch,jonsell,holthaus} and now enjoys
a new life due to its relevance to the experiments discussed
in Ref.~\onlinecite{stof}. In particular, the
theoretical spectrum of this system is in excellent agreement with the experimental spectrum we have
described.\cite{busch}
Our approach
does not make any reference to the shape of the short-range two-body
potential, and requires only that its range be much smaller than the
oscillator length,~\cite{busch} which serves as a natural measure of the particle
spacing. This approach beautifully illustrates the notion of 
shape-independence of potentials (for low
energy scattering and binding) that was first developed in the context
of nuclear physics.\cite{bethe} Shape-independence is embodied in the effective range expansion, which is explained and
derived in Ref.~\onlinecite{blatt}. The 
concepts of scattering length and effective range that come in this
expansion are important in the analysis of the
experimental results with ultra-cold atoms, and readers are encouraged
to study the excellent discussion of these topics in Ref.~\onlinecite{bethe}. 
We shall make use of these concepts and show 
that the energy levels of
this system are independent of the shape of the interaction potential 
in the limit of zero range. This result is accomplished by noting that,
for a harmonic trap, the center of mass motion 
may be isolated, and the interacting relative motion may be written as
a one-body problem. The irregular solution of the harmonic oscillator wave
function, which is normally discarded in the noninteracting problem, 
will be shown to be the relevant solution for the zero-range interacting case.
This realization paves the way for a gentle introduction to the
concept of a pseudopotential in the form of a regularized delta
function.

The zero-range pseudopotential, first introduced by 
Fermi,\cite{fermi} and applied to the many-body problem by Huang and 
Yang,\cite{huang} is extensively used in the current literature
of ultra-cold gases.\cite{hammer} 
The same pseudopotential has also been used to study the spectrum of two 
particles interacting by a zero-range potential in a spherical
harmonic oscillator well.\cite{busch} 
Rather than taking this path, we shall follow the derivation of 
Jonsell,\cite{jonsell} which is in the same spirit as our initial intuitive 
approach. In Sec.~III we finish with some concluding remarks and
suggestions for future research in this area.

\section{The two-body problem}
\subsection{The zero-range pseudopotential}
We first consider the problem of two identical bosons, each
of mass $M$, in the absence of a confining potential. The bosons
interact with a short-range central potential. Most of what follows in
this section also applies to fermionic atoms in the spin-singlet
state, or to nucleons in the $s$-state with an antisymmetric spin-isospin wave
function.\cite{preston} In the relative co-ordinate $r=|{\bf r_1}-{\bf r_2}|$
the $s$-state asymptotic scattering wave function for positive energy
$E$ is given by 
\beq
u(r)=r\psi(r) \sim \sin(kr+\delta(k)), \quad (r> b). 
\label{phase}
\eeq
We have $E=\hbar^2k^2/M$, and 
$b$ is the range of the potential. The effective range
expansion, relating the phase shift to the scattering length $a$ and
the effective range $r_0$ is given by\cite{noyes} 
\beq
k \cot \delta(k)=-\frac{1}{a}+\frac{1}{2}
r_0k^2-Pr_0^3k^4+Qr_0^5k^6+ \ldots
\label{effective}
\eeq
Potentials of different shapes may have identical values of $a$ and
$r_0$, but the parameters $P$ and $Q$ are shape-dependent. Figure 1
schematically illustrates the zero-energy wave function $u(r)$ 
for an attractive square-well potential, 
\begin{figure}[h!]
\centering
\includegraphics[angle=90,height=12cm,width=10cm]{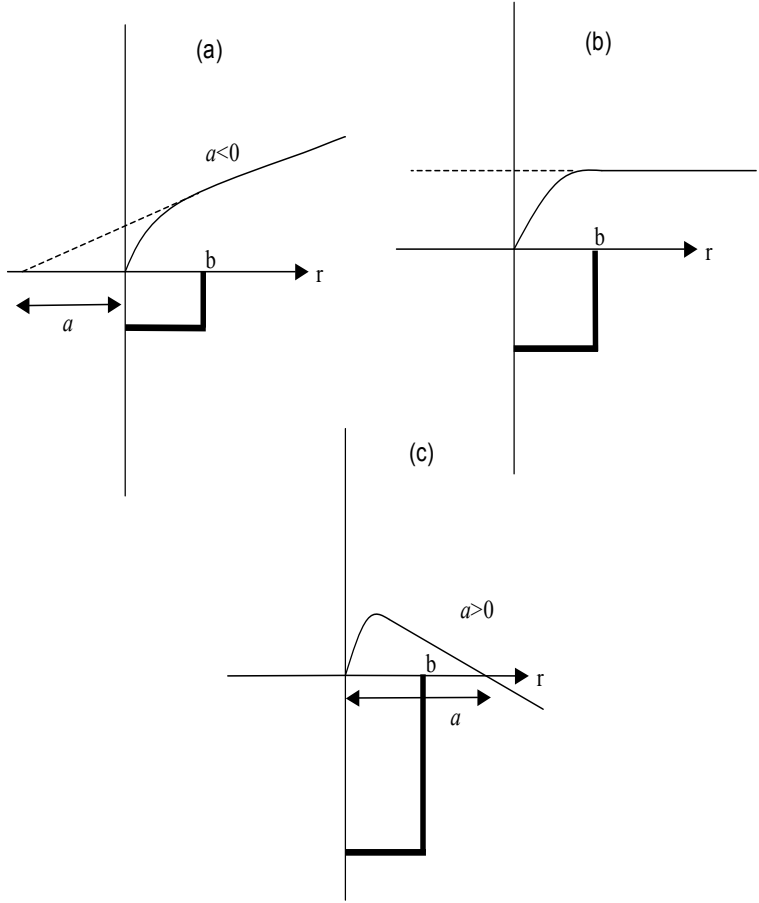}
\caption{An attractive spherical square well potential. The thin solid
line indicates the zero-energy wave function $u(r)$, and the dashed line
the extrapolated asymptote. (a) Shallow well with negative scattering
length. (b) Depth tuned to $V_0 = \pi^2\hbar^2/4Mb^2$ with 
$a\rightarrow \infty$ corresponding to the zero energy bound sate.
(c) A deeper well supporting only one bound state and positive
scattering length. Note that $u(r) \sim (r-a)$ for $r>b$.}
\end{figure}
\beq
V(r)=-V_0\Theta(b-r), 
\label{square}
\eeq
of depth $V_0$ and range $b$; $\Theta(x)$ is the usual unit Heaviside function.

Figure~1 depicts a situation where the intercept of the the 
zero-energy wave function (or its extrapolation) on the horizontal
axis is the scattering length $a$, with its sign signifying the
presence or absence of a bound state. 
For a shallow potential that supports no bound state, the
scattering length $a$ is negative (Fig.~1(a)). When the depth of the
potential is increased and fine-tuned as in Fig.~1(b), the system has a 
zero-energy ``bound'' state corresponding to $a \rightarrow \pm \infty$. For this
limiting case the dimensionless quantity
$\gamma=MV_0b^2/\hbar^2$ equals $\pi^2/4$. The
corresponding values of $\gamma$ for other two-parameter attractive potentials
are listed in Ref.~\onlinecite{pla}.
A further increase in the depth results in supporting only one bound
state, and the scattering length $a$ is shown to be positive
(Fig.~1(c)).

The situation is more complicated when the depth of the potential becomes
still stronger to support two or more bound states.   For the case of the 
square-well example, it may be shown that the $s$-wave scattering length is given by\cite{preston} 
\beq
\label{eq:above}
a=b\left(1-\frac{\tan\sqrt{\gamma}}{\sqrt{\gamma}}\right).
\eeq
Therefore $\sqrt{\gamma}=3\pi/2$ signals the appearance of the next
zero-energy bound state, with the scattering length changing sign. 
A further increase results in two bound
states. Note from Eq.~\eqref{eq:above} that the scattering length $a$ depends on the range
$b$ of the potential and its strength $\gamma$. 
In what follows we shall be interested in
taking the zero-range limit of the interaction, while keeping the
strength parameter $\gamma$ fixed.

The effective range $r_0$ is
proportional to the range $b$ of the potential, and goes to zero as 
$b\rightarrow 0$. 
For this case Eq.~(\ref{effective}) reduces to 
\beq
k \cot \delta=-\frac{1}{a}.
\label{ex}
\eeq
For positive $a$ we
may extrapolate Eq.~\eqref{ex} to a bound state. Note that the
S-matrix for the $s$-state is defined as $S(k)=\exp(2i\delta(k))$, and
the S-matrix has a simple pole for each bound state in the upper half of
the imaginary $k$-axis on the complex $k$-plane.\cite{gottfried} 
A zero-range potential
can support only one bound state. At a bound state therefore, 
$k=i\kappa$, and $\cot \delta=i$. The relation
(\ref{ex}) for a zero-range potential then reduces to $\kappa=1/a$. 
Hence the binding energy $B=-E$, and the bound state wave function
$\psi(r)$ is given by 
\begin{align}
\psi(r)&=\frac{\exp(-r/a)}{r}, \quad (r>0), \label{scat} \\
B & =\hbar^2/Ma^2. \label{binding}
\end{align}
Equation~(\ref{scat}) is exact for a zero-range potential. 
If $r_0$ is nonzero, the relations for $B$ and the asymptotic ($r>b$) 
wave function $\psi$ are approximately
true only if the binding $B$ is close to zero, which implies that
$a \gg r_0$, and the system is large in size compared to the range of the 
potential. As an example where these conditions are only marginally
satisfied, consider the deuteron, which is the lightest stable atomic
nucleus, and is a bound state of a neutron and a proton. The n-p
scattering length in the $s$-state ($S=1$, $T=0$) is $a=5.43 $\,fm, and
$r_0=1.75$\,fm, where $1\,\mbox{fm}=10^{-15}$\,m. Equation~(\ref{binding}) 
for the binding energy gives $B=1.41$\,Mev, compared
with the experimental value of $2.23$ Mev.\cite{round}

Because $r_0$ is
nonzero but is much smaller than $a$, this agreement could be
improved by taking the next term in the expansion in Eq.~(\ref{effective}). 
Following the same extrapolation for the bound state at $k=i\kappa$, 
we obtain the quadratic equation 
\beq
\kappa=\frac{1}{a}+\frac{1}{2} r_0 \kappa^2.
\eeq
We select only the root that gives $\kappa=0$ when
$a\rightarrow\infty$ and $r_0=0$ and obtain 
\beq
\label{formula}
\kappa=\frac{1}{r_0}\left(1-\sqrt{(1-2r_0/a)}\right).
\eeq 
Equation~\eqref{formula} for the deuteron gives $B=2.16$\,MeV, which is close
to the experimental value.\cite{tensor} 

We have assumed that the size of the bound two-body system is much larger
than the range of the potential responsible for the binding. The
details of the potential at short distances do not manifest themselves
significantly in the wave function, and the binding is determined by
the low energy scattering parameters. It should therefore
be possible to design an effective potential that reproduces the
shape-independent results we have obtained. Such an effective 
one-parameter pseudopotential is easily derived using the wave
function $\psi(r)$ given by Eq.~(\ref{scat}).
From the relation 
$\nabla^2(1/r)=-4\pi\delta^3(r)$ we obtain $-\nabla^2(\exp(- r/a)/r)
=-a^{-2} \exp(- r/a)/r + 4\pi \exp(- r/a) \delta^3(r)$. 
If we identify $\exp(- r/a)/r=\psi(r)$, we may write 
\beq
-\nabla^2\psi+4\pi a \delta^3(r)\frac{\partial}{\partial r}
r\psi=-\frac{1}{a^2}\psi.
\label{point}
\eeq
We see that the pseudopotential
\beq
V(r)=\frac{\hbar^2}{M}4\pi a \delta^3(r)\frac{\partial}{\partial r} r
\label{ap2}
\eeq 
reproduces the earlier
results for the wave function and the binding energy of any zero-range 
potential with a given scattering length. In the many-body
problem of an ultra-cold dilute gas, where $k\rightarrow 0$, and the 
average distance between the
particles is much larger than the range $b$ of the potential, the same 
regularized pseudopotential in Eq.~\eqref{ap2}
is widely used.\cite{huang, hammer} Note that $\partial (r\psi)/\partial r$ in the limit $r\rightarrow 0$ is $\psi(0)$ 
for a wave function that is well-behaved at the origin. In addition, it 
also gives a finite result for the irregular $\psi(r)$ that goes as 
$r^{-1}$, which is more relevant for our problem.

\subsection{Two particles in a harmonic potential}
First consider the case of noninteracting particles. Each particle of mass $M$
moves in a harmonic potential $1/2 M\omega^2 r^2$. If we make the usual 
transformations to center-of-mass and relative co-ordinates, ${\bf r}=({\bf r}_1-{\bf r}_2)$, and ${\bf R}=({\bf r}_1+{\bf r}_2)/2$, and their
corresponding canonical momenta ${\bf p}$ and ${\bf P}$, we obtain 
\beq
H_0=(\frac{P^2}{2M_{\rm cm}}+\frac{1}{2} M_{\rm cm}\omega^2R^2) + (\frac{p^2}
{2\mu}+\frac{1}{2} \mu \omega^2 r^2), 
\eeq
where $M_{\rm cm}=2M$ and $\mu=M/2$. 
We concentrate on the relative motion and divide the relative 
distance $r$ by 
$\sqrt{\hbar/\mu \omega}$ and the energy $E$ by $\hbar\omega$.
In terms of the dimensionless scaled variables $x=r/\sqrt{2}l$, 
where $l=\sqrt{\hbar/(M\omega)}$, and $\eta=2E/(\hbar\omega)$, the 
Schr\"odinger equation 
for the relative wave function $u(r)=r\psi(r)$ in the $s$-state is given by 
\beq 
-\frac{d^2 u}{dx^2}+x^2 u=\eta u.
\label{wave}
\eeq
The regular ground state solution is $u_0(x)=x \exp(-x^2/2)$,
corresponding to $\psi(x)=\exp(-x^2/2)$, with $\eta=3$, or $E=\frac{3}{2}
\hbar\omega$. It is easily checked that Eq.~\eqref{wave} is also satisfied for $x>0$ by $u(x)=\exp(-x^2/2)$, corresponding to 
$\psi(x)=\exp(-x^2/2)/x$ with $\eta=1$ or $E=\frac{1}{2} \hbar \omega$.
Although normalizable and lower in energy, it is excluded as a valid
solution in the noninteracting problem because at $r=0$, the kinetic operator 
$\nabla^2$ acting on $1/r$ would yield a delta function. 

The situation changes for the zero-range interaction that we have been 
considering. At $r=0^{+}$ we may ignore the harmonic confinement, and the 
wave function is given by Eq.~(\ref{scat}). Even for nonzero
$b$ the harmonic potential may be ignored if 
$\mu\omega^2b^2 \ll \hbar\omega$, which yields the condition 
$b/l \ll 1$ for the validity of the spectrum derived in the following.
For the scattering length 
$a\rightarrow \infty$, the wave function is $1/r$, and smoothly
joins with the irregular wave function above for $\eta = 1$ This analytic behavior suggests that the
valid (normalized) ground state wave function in the limit 
of $a\rightarrow \infty$ for $r>0$ is the irregular solution, 
\beq
\psi(x)=\sqrt{\frac{2}{\pi^{3/2}}} \frac{\exp(-x^2/2)}{x},
\label{gs}
\eeq
with $E=\frac12\hbar \omega$. The next excited $\ell=0$ state has one
node, and by demanding that it be orthogonal to the the ground state
given by Eq.~(\ref{gs}), may be deduced to be $(x^2-1/2)\exp(-x^2/2)/x$,
with energy $E=\frac52 \hbar \omega$. There is an infinite tower of
such radially excited states with $\ell=0$. This result is confirmed by our 
numerical calculations with the two-body potential 
\beq
V(r)=-V_0{\rm sech}^2(r/b).
\eeq
This potential has a bound state at $E=0$ in free space when 
$V_0=8 \hbar^2/Mr_0^2$, and $b=r_0/2$. With these
parameters, the scattering length $a$ is infinite, and the 
effective range is $r_0$. Figure 2 displays the first five
eigenvalues as a function of $r_0/l$. Note that as $r_0\rightarrow
0$, $E$ in units of $\hbar\omega$ approaches $1/2, 5/2, 9/2, \ldots$. 

\begin{figure}[h!]
\includegraphics[width=\textwidth]{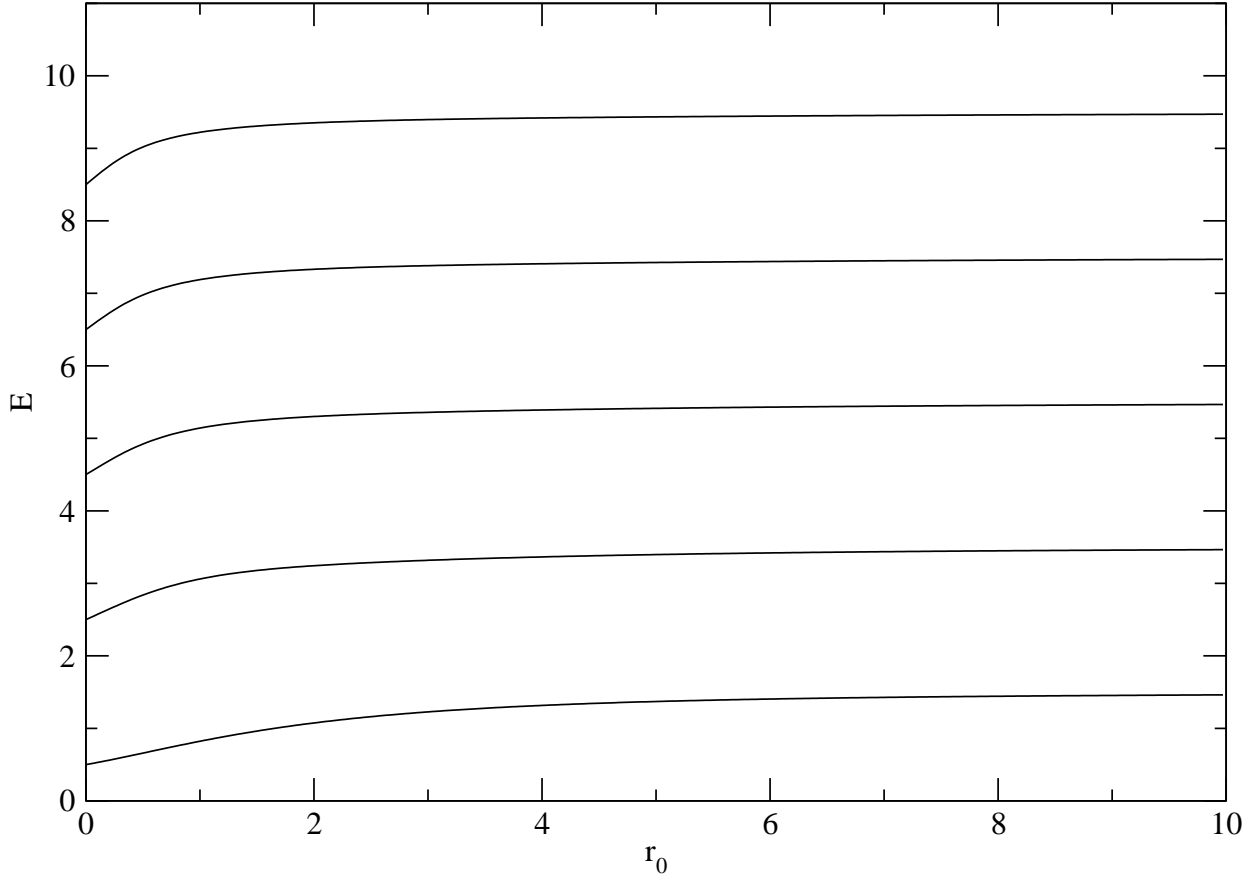}
\caption{The lowest five $\ell = 0$ energy levels for the potential
$V(r) = -(8\hbar^2/Mr_0^2){\rm sech}^2\left(\frac{2r}{r_0}\right)$,
as $r_0$ is decreased toward zero. In this and all subsequent figures,
lengths and energies are scaled as described in the text.}
\end{figure}

The tower of states described above, for $a\rightarrow\infty$, may be generalized 
for any given $a$, following Ref.~\onlinecite{jonsell}. From
our previous considerations it is reasonable to assume that the general
solution for $x>0$ is of the form $u(x)=\exp(-x^2/2) f(x)$. It is
convenient to let $y=x^2$, and $w(y)=f(x)$. After some algebra,
Eq.~(\ref{wave}) is transformed to 
\beq
\label{trans}
y \frac{d^2w}{dy^2} +(\frac{1}{2}-y)\frac{dw}{dy} +(\frac{\eta-1}{4})
w=0.
\eeq 
Equation~\eqref{trans} is the equation satisfied by the confluent
hypergeometric function, and the solution is given by\cite{LL} 
$w(y)=c_1 M((1-\eta)/4, 1/2, y) + c_2 y^{1/2} M((3-\eta)/4, 3/2, y)$, 
where $c_1$ and $c_2$ are constants. We transform back to the $x$
co-ordinate and obtain 
\beq
u(x) =\left[c_1 M((1-\eta)/4, 1/2, x^2) + c_2 x M((3-\eta)/4, 3/2, x^2)\right]
\exp(-x^2/2).
\label{hyper}
\eeq
The ratio $c_1/c_2$ may be
determined by noting that for large $y$,
$M(p,q,y)\sim\Gamma(q)/\Gamma(p) y^{(p-q)}\exp(y)$. 
The large $x$ behavior of the 
solution $u(x)$ given in Eq.~(\ref{hyper}) is
\beq 
u(x)\sim \left[c_1 \frac{\Gamma(1/2)}{\Gamma((1-\eta)/4)}+c_2
\frac{\Gamma(3/2)}{\Gamma((3-\eta)/4)}\right]
x^{-(1+\eta)/2}\exp(x^2/2).
\eeq 
For $u(x)$ not to diverge for large $x$, the quantity in the 
square brackets must vanish, yielding the ratio to be 
\beq
\frac{c_1}{c_2}=-\frac{1}{2}
\frac{\Gamma((1-\eta)/4)}{\Gamma((3-\eta)/4)}.
\label{ratio}
\eeq

We now relate this ratio to the scattering length $a$ to
determine the eigenvalue $\eta$. We already noted 
that as
$r \rightarrow 0^{+}$, the oscillator potential goes to zero, and 
$u(r)$ for positive energy is a scattering solution
outside the zero-range potential, given by Eq.~(\ref{phase}). For small $r$ we have to within an overall constant, 
\beq
u(r)\simeq (r+\frac{1}{k} \tan \delta). 
\label{tan}
\eeq
In contrast, because the confluent hypergeometric functions go to unity
for $x\rightarrow 0^{+}$, we find from Eq.~(\ref{hyper}), 
$u(x)\sim (c_1+c_2 x)$. If we factor $c_2$ out of this expression and rewrite
$u$ in terms of the variable $r$, we find to within an overall constant, 
\beq
u(r)\simeq \Big(r+\sqrt{2}l \frac{c_1}{c_2} \Big).
\label{alt}
\eeq
We compare Eqs.~(\ref{tan}) and (\ref{alt}) and obtain
\beq
\label{above3}
\frac{1}{k} \tan \delta=\sqrt{2}~l \frac{c_1}{c_2}.
\eeq
We substitute $\tan \delta=-a k$ in Eq.~\eqref{above3} and the ratio $c_1/c_2$ from Eq.~(\ref{ratio}) and 
obtain the desired relation 
\beq
\frac{a}{l}=\frac{1}{\sqrt{2}}
\frac{\Gamma((1-\eta)/4)}{\Gamma((3-\eta)/4)}.
\label{fin}
\eeq
The Gamma function diverges when its
argument is zero or a negative integer. Hence, for $a\rightarrow
\infty$, $\eta=1, 5, 9,\ldots$, confirming our earlier results for both
the energy spectrum and the corresponding wave functions $u(x)$.

Note that Eq.~(\ref{fin}) has been obtained with no mention of the
shape of the potential and is therefore applicable for any short-range two-body potential. This general analysis is in contrast with the 
derivation in Ref.~\onlinecite{busch} in which explicit 
use of the properties of the pseudopotential (\ref{ap2}) was required. 

Figure 3 displays the variation of the eigenvalues $E$ (in units of 
$\hbar\omega$) as a function of $a/l$. 
\begin{figure}[h!]
\includegraphics[width=\textwidth]{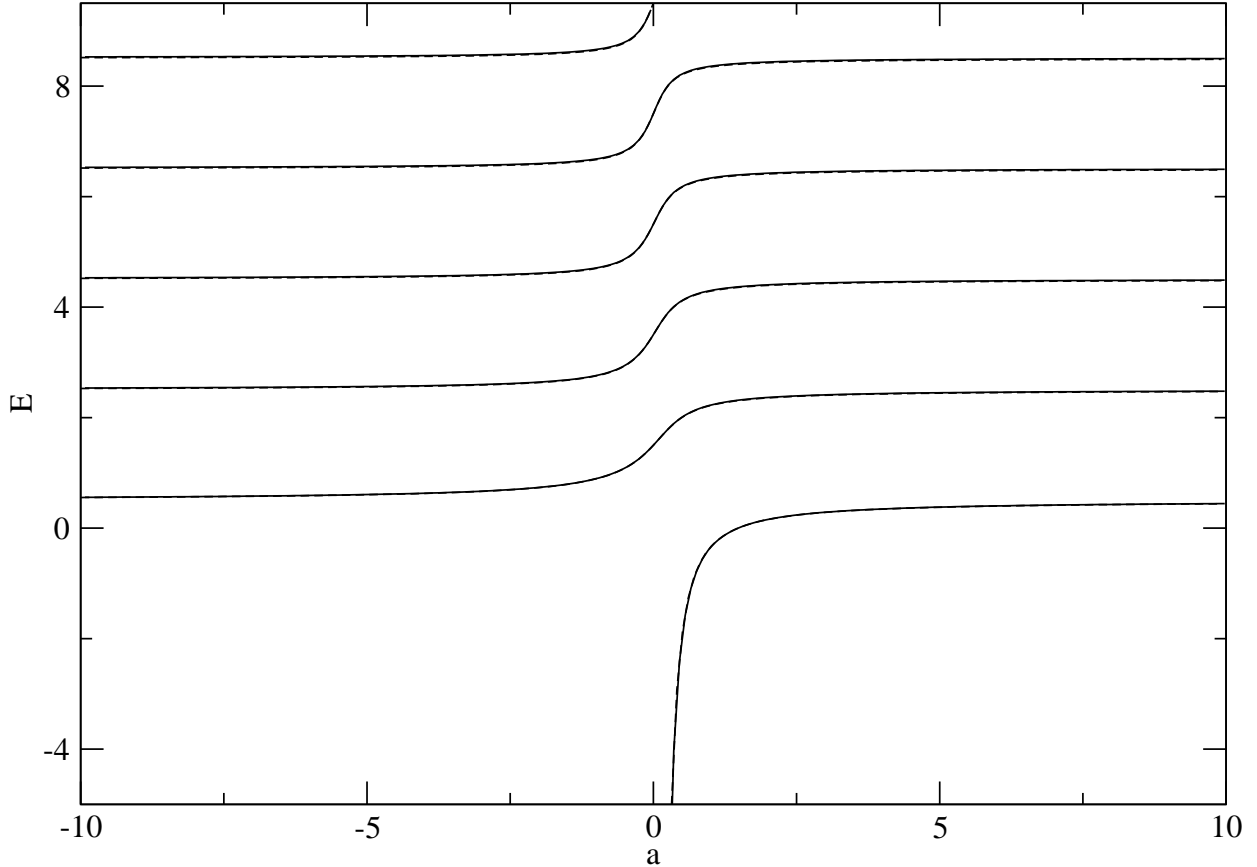}
\caption{The energy spectrum of the two-particle system in a harmonic
trap. The solid lines are for the attractive square-well potential
(\ref{square}) with $b/l = 0.01$, and the dashed lines for the 
pseudopotential (\ref{ap2}). This theoretical spectrum with varying 
scattering length is in excellent agreement with
the experimentally measured binding energies of the dimers formed from
$^{40}$K atoms.\cite{stof}}
\end{figure}
For comparison we also give the values obtained numerically for a square-well potential with $b/l=0.01$. On the scale of the plots, it is
difficult to distinguish between the results. 
Note also that the ground state energy diverges toward $-\infty$ as 
$a\rightarrow 0$. This behavior is the same as that of the $a^{-2}$ divergence of the single
bound state for the regularized delta function potential given by 
Eq.~(\ref{scat}). Note
that a zero scattering length does not necessarily mean
that the interaction is zero. If the scattering length approaches
$0$ from the negative side (see Fig.~1(a)), the interaction becomes weaker and eventually vanishes. If it
approaches zero from the positive side (see Fig.~1(c)), the interaction 
becomes stronger and stronger, squeezing the wave function more and more
inside the potential.
The flow of the
eigenvalue curves as a function of $a$ are to be understood with these
observations in mind. Starting from the left in Fig.~3 with
large negative
$a$, the interaction becomes weaker $a$ increases, eventually
reaching zero when $a\rightarrow 0^{-}$. At this value we recover
the unperturbed eigenvalues of the harmonic oscillator. To explain the
behavior of the curves in Fig.~3 for $a>0$, start from the
right-hand end and decrease $a$. We have already commented on the diverging
ground state. The next one, starting at $5/2$ at the right end,
continually becomes less as $a$ decreases, because the interaction becomes
stronger, eventually reaching the lower value $3/2$ at $a=0^{+}$. 
The previous arguments should help to demystify the counterintuitive 
features of the pseudopotential discussed 
in Ref.~\onlinecite{busch}.

\section{Conclusions}
We have discussed the
two-body problem in the context of ultra-cold atoms. We have
shown that the energy spectrum is a universal property of
the interacting two-body problem provided that the range of the
interaction is much smaller than the oscillator length. 
We have clarified the relation between the
irregular solution and the regularized
$\delta$-pseudopotential. We invite the interested reader to apply
the methods used here to the interesting one- and two-dimensional analogues of this
problem. 
The energy spectra for these cases are also given in Ref.~\onlinecite{busch}. 
The two-dimensional case is elaborated in Ref.~\onlinecite{khare}. 
We suggest Ref.~\onlinecite{lipkin} and \onlinecite{adhikari} for background reading on one- and two-dimensional scattering.

\begin{acknowledgements}
We would
like to thank Yi Liang and Don Sprung 
for going through the manuscript carefully, and both of the anonymous
referees for their very useful suggestions. This research was financed by grants from NSERC (Canada).
\end{acknowledgements}


\begin{thebibliography}{99}

\bibitem{greiner} M. Greiner, O. Mandel, T. Esslinger, T. W. H\"ansch and I. Bloch, ``Quantum phase transition from a superfluid to a Mott insulator in a gas of ultracold atoms,''Nature (London) {\bf 415}, 39--44 (2002). 

\bibitem{stof} T. St\"oferle, H. Moritz, K. G\"unter, M. Kh\"ol, and T. Esslinger, ``Molecules of fermionic atoms in an optical laser,'' Phys. Rev. Lett. {\bf 96}, 030401-1--030401-4 (2006). 

\bibitem{feshbach} H. Feshbach, ``A unified theory of nuclear reactions, II,'' Ann. Phys. (NY) {\bf 281}, 519--546 (2000).

\bibitem{ino} S. Inouye, M. R. Andrews, J. Stenger, H.-J Miesner, D. M. Stamper-Kurn, and W. Ketterle , ``Observation of Feshbach resonance in Bose-Einstein Condensates,'' Nature (London) {\bf 392}, 151--154 (1998); Ph. Courteille, R. S. Freeland, D. J. Heinzen, F. A. van Abeelen, and B. J. Verhaar, ``Observation of Feshbach resonance in cold atom scattering,'' Phys. Rev. Lett. {\bf 81}, 69--72 (1998). 

\bibitem{tannoudji} A clear description of the underlying physics of Feshbach
resonance is given in Claude Cohen-Tannoudji Lectures on
``Fundamental Concepts for Atomic Gases," in {\it Lectures on Quantum Gases}, 
Institut Henri Poincar\'e, Paris, April 2007, (unpublished). PDF available at 
http://www.phys.ens.fr/$\sim$ castin/progtot.html)

\bibitem{note1} For a neutral atom the number of protons in the nucleus
equals the number of electrons. So whether the atom as a whole behaves
as a boson or fermion is determined by the neutron number being either
even or odd. $^{40}$K has $21$ neutrons, and so as a whole behaves
like a fermion.

\bibitem{busch} T. Busch, B.-G. Englert, K. Rzazewski, and M. Wilkens, ``Two cold atoms in a harmonic trap," Found. Phys. {\bf 28}, 549--559 (1998).

\bibitem{jonsell} S. Jonsell, ``Interaction energy of two trapped bosons with long scattering lengths," Few-body Systems {\bf 31}, 255--260 (2002).

\bibitem{holthaus} M. Block and M. Holthaus, ``Pseudopotential approximation in a harmonic trap," Phys. Rev. A{\bf 65}, 052102-1--4 (2002).

\bibitem{bethe} H. A. Bethe, ``Theory of the effective range in nuclear scattering," Phys. Rev. {\bf 76}, 38--50 (1949).

\bibitem{blatt} J. M. Blatt and V. F. Weisskopf, {\it Theoretical
Nuclear Physics} (John Wiley \& Sons, NY, 1952), pp. 56--63.

\bibitem{fermi} An account of the delta-function potential introduced by Fermi is in R. G. Sachs, {\it Nuclear Theory} (Addison-Wesley, Reading, MA, 1953), Chap. 5, p. 89.

\bibitem{huang} K. Huang, {\it Statistical Mechanics} (John Wiley \& Sons, NY, 1965), p. 275; K. Huang and C. N. Yang, Phys. Rev. {\bf 105}, 767--775 (1957).

\bibitem{hammer} Eric Braaten and H. W. Hammer, ``Universality in 
few-body systems with large scattering length," 
Phys. Rept. {\bf 428}, 259--390 (2006). 

\bibitem{preston} See, for example, M. A. Preston and R. K. Bhaduri, 
{\it Structure of the Nucleus} (Addison-Wesley, Reading, MA, 1975), p. 50, and Problem 2-3, p. 61.

\bibitem{noyes} H. P. Noyes, ``The nucleon-nucleon effective range 
expansion parameters," Ann. Rev. Nucl. Sci. 465--484 (1972).

\bibitem{pla} R. K. Bhaduri, M. Seidl, and A. Khare, ``An approximate shape-independent relation at Feshbach resonance," Phys. Lett A {\bf 372}, 4160--4163 (2008).

\bibitem{gottfried} K. Gottfried, {\it Quantum Mechanics} (W. A. Benjamin, New York, 1966), Vol. 1, p. 126.

\bibitem{round} The values given for $a$, $r_0$, and $B$ are given to two decimals for simplicity. We have taken 
$\hbar^2/M=41.5$\,MeV\,fm$^2$, where an average of the neutron and proton
masses have been taken.

\bibitem{tensor} A more careful treatment would include the small admixture of the $d$ state in the deuteron due to the non-central 
tensor force.

\bibitem{LL} L. D. Landau and E. M. Lifshitz, {\it Quantum Mechanics} 
(Addison-Wesley, Reading, MA, 1958), p. 496.

\bibitem{khare} M. Combescure, A. Khare, A. K. Raina, J.-M Richard, and C. Weydert, ``Level arrangements in exotic atoms and quantum dots,'' Int. J. Mod. Phys. B {\bf 21}, 3765--3781
(2007). 

\bibitem{lipkin} H. J. Lipkin, {\it Quantum Mechanics} ( North-Holland, Amsterdam, 1973), p. 199.

\bibitem{adhikari} S. K. Adhikari, ``Quantum scattering in two dimensions,'' Am. J. Phys. {\bf 54}, 362--367 (1986). 

\end{thebibliography}
\end{document}